# Inferring the *in vivo* looping properties of DNA

Leonor Saiz, J. Miguel Rubi[*], and Jose M. G. Vilar

*Integrative Biological Modeling Laboratory, Computational Biology Program, Memorial Sloan-Kettering Cancer Center, 307 East 63rd Street, New York, NY 10021, USA*

[*]*On leave from: Departament de Fisica Fonamental, Universitat de Barcelona, Diagonal 647, 08028 Barcelona, Spain*

**Classification:** BIOLOGICAL SCIENCES: Biophysics

**Corresponding author:**

Jose M.G. Vilar

Computational Biology Center

Memorial Sloan-Kettering Cancer Center

307 East 63rd Street

New York, NY 10021, USA

Tel: 646-735-8085

Fax: 646-735-0021

E-mail: vilar@cbio.mskcc.org

**Manuscript information:** 16 text pages and 4 Figures.

**Word and character counts:** Abstract: 120 words. Total number of characters in paper: 26533 (18097 characters for text + 7596 characters for figures + 720 characters for allowance + 120 characters for equations)

**ABSTRACT**

**The free energy of looping DNA by proteins and protein complexes determines to what extent distal DNA sites can affect each other. We inferred its *in vivo* value through a combined computational-experimental approach for different lengths of the loop and found that, in addition to the intrinsic periodicity of the DNA double helix, the free energy has an oscillatory component with about half the helical period. Moreover, the oscillations have such an amplitude that the effects of regulatory molecules become strongly dependent on their precise DNA positioning and yet easily tunable by their cooperative interactions. These unexpected results can confer to the physical properties of DNA a more prominent role at shaping the properties of gene regulation than previously thought.**

# Introduction

The cell is a densely packed dynamic structure made of thousands of different molecular species that orchestrate their interactions to form a functional unit. Such complexity poses a strong barrier for experimentally characterizing the cellular components: not only the properties of the components can change when studied *in vitro* outside the cell but also the *in vivo* probing of the cell can perturb the process under study (1). Here we use computational modeling to obtain the properties of the *in vivo* unperturbed components at the molecular level from physiological measurements at the cellular level. Explicitly, we infer the *in vivo* free energies of DNA looping from enzyme production in the *lac* operon (2).

The formation of DNA loops by the binding of proteins at distal DNA sites plays a fundamental role in many cellular processes, such as transcription, recombination, and replication (3-5). In gene regulation, proteins bound far away from the genes they regulate can be brought to the initiation of transcription region by looping the intervening DNA. The free energy cost of this process determines how easily DNA can loop and therefore the extent to which distal DNA sites can affect each other (5).

In the *lac* operon, there is a repressor molecule that regulates transcription by binding specifically to DNA sites known as operators and preventing the RNA polymerase from transcribing the genes. DNA looping allows the repressor to bind to two operators simultaneously, leading to an increase in repression of transcription. This increase, characterized by the repression level, can be connected to the free energy of looping DNA by a recent model for transcription regulation by the *lac* repressor (6). The distance between operators determines the length of the DNA loop and affects the repression level through the changes in the free energy of looping. For inter-operator distances from 57.5 bp to 98.5 bp, Muller et al. (7) systematically varied the distance between two operators in increments of 1 bp and measured the *in vivo* repression levels under conditions similar to wild-type. These physiological measurements of enzyme production in *E. Coli* cell populations allowed us to infer here the *in vivo* looping properties of DNA with unprecedented accuracy.

## Background and Methods

**Transcription regulation in the *lac* operon**

The *lac* operon consists of a regulatory domain and three genes required for the uptake and catabolism of lactose. Binding of the *lac* repressor to the main operator $O1$ prevents the RNA polymerase from binding to the promoter and transcribing the genes. There are also two auxiliary operators outside the regulatory region, $O2$ and $O3$, to which the repressor can bind without preventing transcription. They enhance repression of transcription by increasing the ability of the repressor to bind $O1$. This effect is mediated by DNA looping, which allows the simultaneous binding of the repressor to the main and one auxiliary operator (2).

To avoid the formation of multiple loops, experiments are typically performed with just the main and one auxiliary operator, as in Muller et al. (7). In such a case, there is transcription when the system in a state in which:

    (i)    none of the operators is occupied,

    (ii)    a repressor is bound to just the auxiliary operator,

and there is no transcription when:

    (iii)    a repressor is bound to just the main operator,

    (iv)    a repressor is bound to both the main and the auxiliary operators by looping the intervening DNA, and

    (v)    one repressor is bound to the main operator and another repressor, to the auxiliary operator.

In Fig. 1, we illustrate these five possible states for the *lac* constructs used in the experiments of Ref. (7), which consisted of the main operator $O1$ and just an upstream auxiliary operator, $Oid$, with the sequence of the ideal operator.

**Repression level**

The repression level, $R_{loop}$, is a dimensionless quantity used to measure the extent of repression of a gene. It is defined as the ratio of the unrepressed transcription rate ($t_{max}$) to the actual transcription rate ($t_{act}$): $R_{loop} = t_{max}/t_{act}$. The actual transcription rate is the unrepressed transcription rate times the fraction of time that the main operator is free. Therefore, the repression level is the inverse of the probability that the system is in the states (i) and (ii):

$$R_{loop} = \frac{1}{P_i + P_{ii}},$$

where $P$ indicates the probability for the system to be in the state denoted by its subscript.

## Statistical thermodynamics analysis

Statistical thermodynamics connects the probability $P_k$ of a state $k$ with its standard free energy $\Delta G_k$ through $P_k = \frac{1}{Z}[N]^{n_k} e^{-\Delta G_k/RT}$, where $[N]$ is the concentration of repressors expressed in moles; $n_k$ is the number of repressors considered in the state $k$; $RT$ is the gas constant, $R$, times the absolute temperature, $T$; and $Z = \sum_k P_k$ is the normalization factor (8).

This relationship between probabilities and free energies leads to
$$R_{loop} = 1 + \frac{[N]e^{-\Delta G_{O1}/RT} + [N]e^{-(\Delta G_{O1}+\Delta G_{Oid}+\Delta G_l)/RT} + [N]^2 e^{-(\Delta G_{O1}+\Delta G_{Oid})/RT}}{1+[N]e^{-\Delta G_{Oid}/RT}},$$
where $\Delta G_{O1}$ and $\Delta G_{Oid}$ are the standard free energies of binding of the repressor to the $O1$ and $Oid$ operators, respectively, and $\Delta G_l$ is the free energy of looping (6).

The free energy of looping is defined from the free energy of the looped state (iv), $\Delta G_{iv}$, which can be decomposed in binding to $O1$ and $Oid$ and looping contributions: $\Delta G_{iv} = \Delta G_{O1} + \Delta G_{Oid} + \Delta G_l$ (Figure 2). The free energies of the other states are: $\Delta G_i = 0$, $\Delta G_{ii} = \Delta G_{Oid}$, $\Delta G_{iii} = \Delta G_{O1}$, and $\Delta G_v = \Delta G_{O1} + \Delta G_{Oid}$.

For a strong auxiliary operator ($[N]e^{-\Delta G_{Oid}/RT} \gg 1$), as in the experimental conditions analyzed (7), for which $[N]e^{-\Delta G_{Oid}/RT} \approx 135$, the previous expression simplifies to $R_{loop} = 1 + \left(e^{-\Delta G_l/RT} + [N]\right)e^{-\Delta G_{O1}/RT}$. When the auxiliary operator is not present ($\Delta G_l = \text{infinity}$), the repression level is given by $R_{noloop} = 1 + [N]e^{-\Delta G_{O1}/RT}$.

77

## Results and Discussion

The mathematical expressions connecting the repression level with the molecular properties for a strong auxiliary operator, $R_{loop} = 1 + \left(e^{-\Delta G_l/RT} + [N]\right)e^{-\Delta G_{O1}/RT}$, and for no DNA looping, $R_{noloop} = 1 + [N]e^{-\Delta G_{O1}/RT}$, have the remarkable property that they can be combined to give the free energy of looping $\Delta G_l$ as a function of the physiologically measurable quantities $R_{loop}$ and $R_{noloop}$:

$$\Delta G_l = -RT \ln \frac{R_{loop} - R_{noloop}}{R_{noloop} - 1} [N]. \tag{1}$$

We used this expression with measured repression levels (7) (Figure 3a) to obtain the *in vivo* free energies of looping DNA by the *lac* repressor for different distances between operators (Figure 3b). This mathematical transformation reveals a wealth of details that were not evident in the repression level curves (7, 9). As expected, the free energy oscillates with the helical periodicity of DNA (7, 9-11) since the operators must have the right phase to bind simultaneously to the repressor. The free energy in a cycle, however, behaves asymmetrically, increasing much faster (~3 bp) than decreasing (~8 bp) as the length of the loop increases. A Fourier analysis of the oscillations (Figure 3c) indicates that, in addition to the component with the helical period (~10.9 bp), there is a second representative component with period ~5.6 bp, which leads to the observed asymmetry (Figure 3d). This second component has a strong statistical significance (see Appendix I) and is also present for other experimental data (see Appendix II).

Previous *in vivo* experiments and analysis for longer loops (>100 bp) used continuum elastic models of DNA to fit the observed repression levels (12-14). Therefore, the conclusions that can be extracted from those studies are constrained by the properties of the DNA models used, which are unable to account for the asymmetry we observe in the oscillations for short loops (Figure 3b). Our analysis, in contrast, does not depend on any underlying DNA model and is able to capture novel properties, not considered by current DNA looping models. Recent structural and computational studies on DNA (15, 16) indicate that the loop can be bent and twisted non-uniformly because of different contributions, as for instance, the anisotropic flexibility of DNA, local features resulting from the DNA sequence, and interactions with the lac repressor and other DNA binding proteins. Therefore, the behavior we observe might result from the detailed molecular structure of DNA (15, 16) or the intrinsic elasticity of the lac repressor (17).

Another remarkable property of the free energy is that the amplitude of the oscillations is ~2.5 kcal/mol. Such an amplitude, which is much smaller than predicted from current DNA models (15, 18) and similar to the typical free energies of interaction between regulatory molecules (19), demonstrates that the effects of regulatory molecules are strongly dependent on their precise DNA positioning and at the same time easily tunable and modifiable by their cooperative interactions.

Calibrated models, as the one we have used here, are widely used in physics and engineering as measuring tools. A well-known example is the measurement of the temperature from the height of the liquid column of a thermometer. In the prototypical model of a thermometer, the temperature in Celsius is given by $T = \frac{h - h_{freeze}}{h_{boil} - h_{freeze}} 100$, where $h$ is the height of the column and the subscripts boil and freeze label the heights at which water boils and freezes, respectively (20). Our results show that analogous measuring tools can also be implemented in cellular systems to infer molecular properties from the observed physiological behavior. Explicitly, the *lac* operon can function as a sensor to measure *in vivo* free energies of DNA looping. The results we have obtained with this combined computational-experimental approach challenge the universality of current continuum elastic DNA models at short distances and point to an active role of the physical properties of DNA at shaping the properties of gene regulation.

# Appendix I: Statistical reliability of the half-helical-period component

To study the effects of noise in the measurements we consider that the free energy $\Delta G'_l(l)$ for a loop length $l$ is affected by a random quantity $\xi(l)$ with zero mean:

$$\Delta G_l(l) = \Delta G'_l(l) + \xi(l).$$

The Fourier transform of the deterministic and random components at the frequency $v_{1/2}$ of the half-helical-period component are defined as

$$F(\Delta G'_l, v_{1/2}) = \frac{1}{M}\sum_l \Delta G'_l(l)\exp(2\pi i v_{1/2} l) \text{ and } F(\xi, v_{1/2}) = \frac{1}{M}\sum_l \xi(l)\exp(2\pi i v_{1/2} l),$$

respectively, where $M$ is the number of measurements.

The power spectrum is defined as the squared modulus of the Fourier transform. Therefore, the relative contribution of the measurement noise is

$$NSR^2 = \frac{F(\xi, v_{1/2})F(\xi, -v_{1/2})}{F(\Delta G'_l, v_{1/2})F(\Delta G'_l, -v_{1/2})} \approx \frac{\sigma^2}{MA_{1/2}^2},$$

where $\sigma^2 = \frac{1}{M-1}\sum_l \xi(l)^2$ is the variance of the random contribution to the measured free energies and $2A_{1/2}$ is the amplitude of the half-helical-period component.

From the variance of the free energies in Table 2 of reference (6) one can estimate $\sigma = 0.15 \text{ kcal/mol}$. From Muller et al.'s data we obtain $2A_{1/2} = 0.30 \text{ kcal/mol}$ (Figure 3d), $v_{1/2} = 5.6$ bp (Figure 3c), and $M = 41$ (Figure 3a). Therefore, the relative contribution of the measurement noise to the power spectrum at the 5.6 bp frequency is $NSR = 0.16$, which indicates that only 16% of amplitude of the half-helical-period component can be attributed to measurement errors and that the probability of obtaining the 5.6 bp peak of the power spectrum just by chance is effectively zero.

The reliability of the half-helical-period component is much higher than that of single measurements because the evaluation of this component incorporates all the 41 measurements of the free energy for the different loop lengths. For large $M$, approximately $M > 10$, the error can be approximated by a Gaussian distribution and the probability of obtaining a periodic component as high as $A_{1/2}$ because of the measurement errors is given by

$$P(A_{1/2}) = \frac{2}{\sqrt{2\pi\sigma^2/M}} \int_{A_{1/2}}^{\text{infinity}} e^{-M\frac{x^2}{2\sigma^2}} dx .$$

Only for high values of the measurement error does this probability become appreciable. For instance, in the previous case, even if the measurement errors had a standard deviation $\sigma = 0.3$ kcal/mol the probability of obtaining a periodic component with amplitude 0.30 kcal/mol or higher as a result of the measurement noise would be below 0.2% ($P(0.3) = 0.0014$).

## Appendix II: Corroboration of the main results

The main results we have obtained based on the classical experiments of Muller et al (7) are corroborated by the application of our method to the recent experimental data of Becker et al. (21). These new experiments differ from Muller et al.'s mainly in quantitative details. Becker et al. used a weaker main operator, a lower number or repressors per cell, and measured the repression levels for fewer loop lengths. Because all these differences have a tendency to increase the statistical fluctuations, the results obtained with this new data (Figure 4) are expected to be slightly less precise than the ones we obtained in Figure 3.

Similarly as we did in Figure 3, we used Equation 1 with the new measured repression levels (21) (Figure 4a) to obtain the *in vivo* free energies of looping DNA by the *lac* repressor for different distances between operators (Figure 4b). A Fourier analysis of the oscillations (Figure 4c) also reveals the component with the helical period (~11.6 bp) and a second representative component with period ~5.2 bp, which lead to the asymmetric behavior (Figure 4d).

**Figure Legends**

**Figure 1:** Schematic representation of the five possible states for *lac* repressor binding to two operators for the *lac* constructs used in the analyzed experiments by Muller et al. (7). The *lac* Z gene, downstream the main operator, is repressed when the *lac* repressor (in gray with black contour line) is bound to the main operator: states (iii), (iv), and (v); and unrepressed when the main operator is unoccupied: states (i) and (ii). The thick black line represents the DNA and the two *lac* operators are shown as white (auxiliary operator) and highlighted (main operator) boxes. The dotted thick line indicates the different lengths of the spacer DNA between operators.

**Figure 2:** Cartoon of the *lac* repressor (in gray with black contour line) bound simultaneously to the main ($O1$) and one auxiliary operator ($Oid$, with the sequence of the ideal operator), which are represented by white rectangular boxes. Binding of the repressor to $O1$ represses the *lac* Z gene in the experiments (7). The simultaneous binding to $O1$ and $Oid$ with free energies $\Delta G_{O1}$ and $\Delta G_{Oid}$, respectively, competes with the unfavorable process of forming a loop ($\Delta G_l$). Using this type of *lac* constructs, Muller et al. (7) measured the *in vivo* repression of the *lac* Z gene as a function of the distance between the centers of the $O1$ and $Oid$ operators.

**Figure 3:** From enzyme content of *E. coli* populations to *in vivo* DNA looping free energies. (**a**) Measured repression levels, $R_{loop}$, as a function of the distance between operators from Muller et al. (7) (**b**) Inferred free energies of looping using Equation 1 and $R_{loop}$ as in panel **a**, with $R_{noloop} = 135$ and $[N] = 75$ nM (7) (**c**) Fourier analysis of the operator distance dependence of the free energy. (**d**) Behavior of the two main periodic components of the free energy of looping:

$\Delta G_l(l) = 8.79 - 0.68\sin(2\pi l/10.9 - 0.94) - 0.30\sin(2\pi l/5.6 - 2.32)$, where $l$ is the distance between operators.

**Figure 4:** Corroboration of the main results. (**a**) Measured repression levels, $R_{loop}$, as a function of the distance between operators from Figure 4a of Becker et al. (21). (**b**) Inferred free energies of looping using Equation 1 and $R_{loop}$ as in panel **a**, with $R_{noloop} = 5$ and $[N] = 15$ nM. (**c**) Fourier analysis of the operator distance dependence of the free energy. (**d**) Behavior of the two main periodic components of the free energy of looping: $\Delta G_l(l) = 9.58 - 0.72\sin(2\pi l/11.6 - 3.08) - 0.28\sin(2\pi l/5.2 - 3.31)$, where $l$ is the distance between operators.

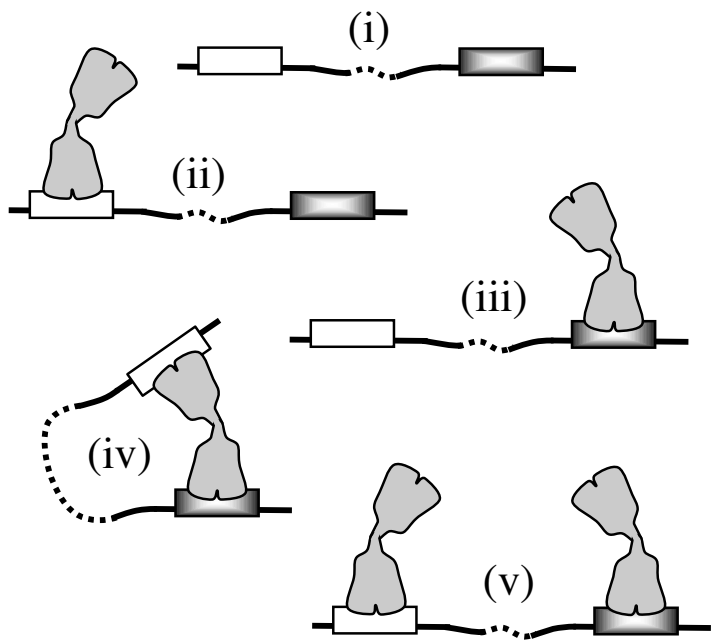

Figure 1. Saiz, Rubi, and Vilar

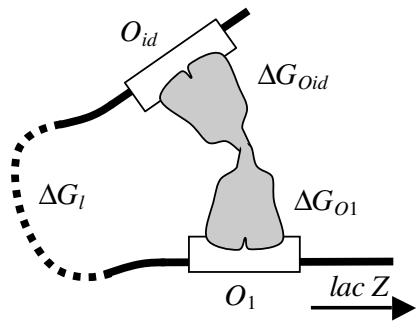

Figure 2. Saiz, Rubi, and Vilar

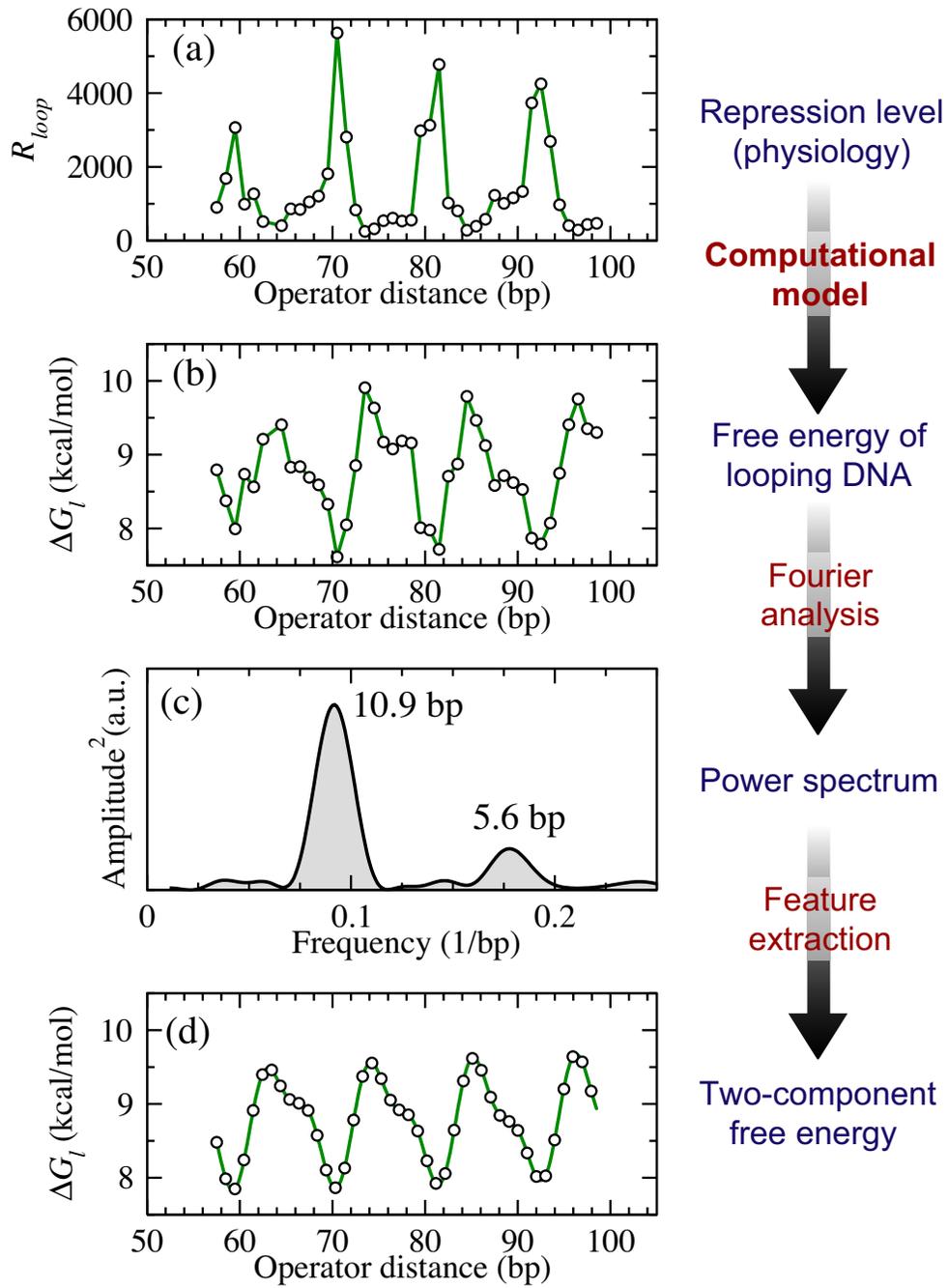

Figure 3. Saiz, Rubi, and Vilar

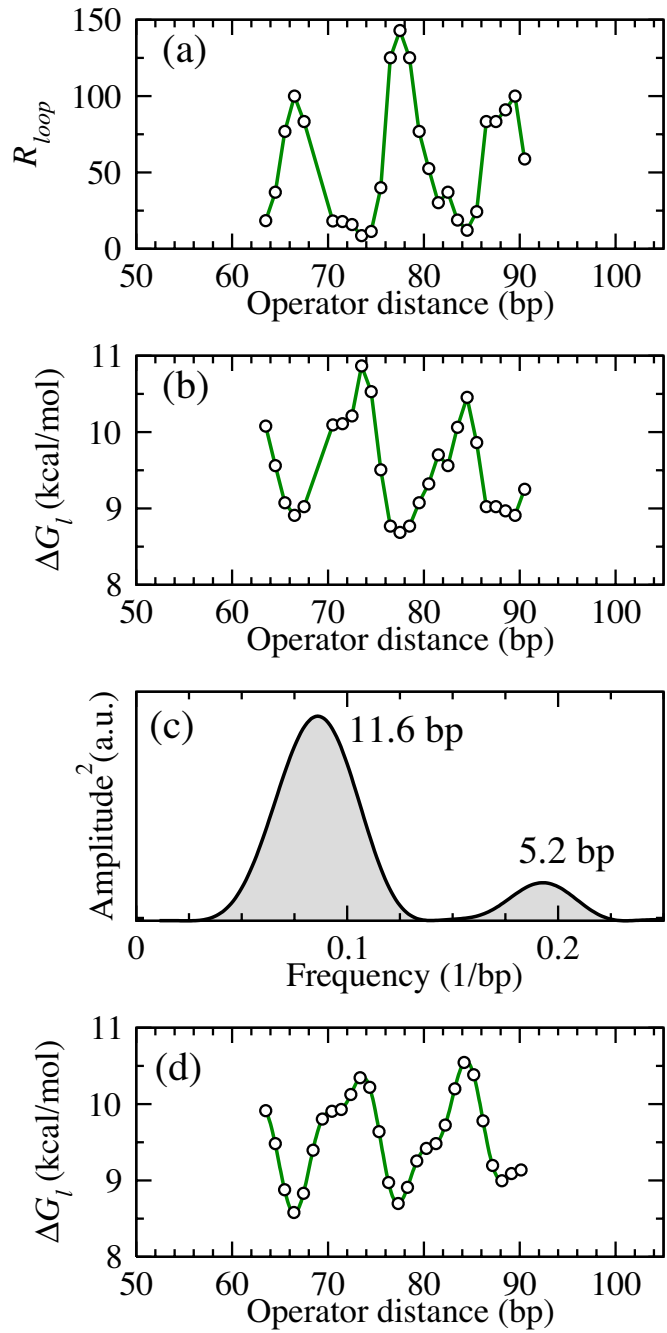

Figure 4. Saiz, Rubi, and Vilar